\documentstyle[12pt]{article}

\input{psfig}

 \newcommand{\bpi}{\mbox{\boldmath $\pi$}}
\title{\begin{flushright}
{\normalsize NUC-MINN-96/3-T\\
February 1996 \\}
\end{flushright}
\vspace*{0.3in}
{\bf PION DECAY CONSTANT AT FINITE TEMPERATURE IN THE NONLINEAR
SIGMA MODEL}}
\author{{\bf Sangyong Jeon} and {\bf Joseph Kapusta}\\
  {\it School of Physics and Astronomy}\\
   {\it University of Minnesota}\\ {\it Minneapolis, MN 55455}}

\date{}

\begin{document}

\maketitle

\begin{center}
Abstract
\end{center}

\noindent
We calculate the pion decay constant near the critical temperature
of the $O(N)$ nonlinear sigma model in the large $N$ limit.
Making use of the known low temperature behavior, we construct
a Pad\'{e} approximant to obtain the behavior of $f_{\pi}(T)$ at
all temperatures.\\

\vfill \eject

According to folklore a nonlinear sigma model, defined by the Lagrangian
\begin{equation}
{\cal L} = - \frac{1}{2} \left( \partial_{\mu}\bpi \right)^2
- \frac{1}{2}\left( \partial_{\mu} \sigma \right)^2  \, ,
\label{eq:lagrangian}
\end{equation}
together with the constraint
\begin{equation}
\bpi^2(x) + \sigma^2(x) = f_\pi^2  \, ,
\label{eq:constraint_micro}
\end{equation}
does not have a phase transition at any finite temperature.
In other words, due to the constraint
the magnitude of the $\sigma$ field is fixed at
the value of $f_\pi$ and hence the symmetry is broken no matter
what the temperature is.  However, this argument is wrong.
A recent analysis \cite{BK} shows explicitly that a symmetry
restoring phase transition occurs at
\begin{equation}
T_c^2 = 12 f_\pi^2/(N + 2)
\end{equation}
to leading plus subleading order in the $1/N$ expansion,
where $N - 1$ is the number of pion field components.

One may ask how it is possible to restore the $O(N)$ symmetry when
the constraint forces the symmetry to be broken by the ground state.
The answer is that a phase transition takes place
not because the symmetry of the ground state changes,
but because the probability of occupying the ground state decreases
as the temperature increases,
becoming infinitesimal at a transition temperature.
For a familiar example, consider a ferromagnet.
The magnitude of the spin at each site is
fixed by the constraint ${\bf S}^2 = S(S{+}1)$.
The ground state of this system is of course one where all the spins line
up in a fixed direction, breaking the rotational symmetry of space.
However, this system certainly goes through a symmetry
restoring phase transition as the temperature is raised.
At high temperatures the amount of energy put into the system is
simply too great to be accommodated by varying the phase of the
ground state alone.

Given that there is a phase transition in the nonlinear $O(N)$ sigma model,
it is important to ask how the pion decay constant changes as the
temperature increases.  One must carefully define what is meant
by the pion decay constant at finite temperature.  A reasonable
definition is that it measures the strength of the coupling of
the pion, the Goldstone boson, to the longitudinal part of the
correlator of two axial vector currents.  Following \cite{BK}
we take as the definition
\begin{equation}
f_\pi^2(T)  =  2 \lim_{\epsilon\to 0}\,
\int_0^{\epsilon} \, {dq_0^2 \over q_0^2}\,
\rho_L(q_0, {\bf q}=0)  \, .
\label{eq:f_pi_def}
\end{equation}
Here $\rho_L(q)$ is the longitudinal part of the spectral density for
the retarded response function of two axial vector currents, where
\begin{equation}
{\cal A}_\mu^a  =  \sigma \partial_\mu\pi_a - \pi_a\partial_\mu\sigma
\end{equation}
in the sigma model.  To give a more physical meaning,
consider first what pions are in a finite temperature system.
Pions are Goldstone bosons associated with the broken chiral
symmetry, and in a finite temperature system represent the energy
stored in a varying phase structure of the ground state.
Consequently, the pion contribution to the correlation function
must depend on the probability of the system to be in the ground state.
That probability is relatively large at low temperatures
and so the pion decay constant is not much different from the zero
temperature value.  As the temperature goes up, the probability of
being in the ground state diminishes, and so does the pion decay constant.
At the phase transition temperature no pions are present in the
system since no energy is stored in the varying phase of the ground state,
and the pion decay constant as well as the scalar condensate
becomes zero.  This all assumes the transition is second order, of course.

The constraint is local and difficult to deal with directly.
It is advantageous to represent it with a Lagrange multiplier $m^2$ and
an auxiliary field $\lambda(x)$
as is done in \cite{Polyakov,BK}.  In Euclidean space,
used throughout this paper,
\begin{equation}
{\cal L} = - \frac{1}{2}\left[ (\partial_{\mu}\bpi)^2 +
(\partial_{\mu} \sigma)^2 + m^2 (\bpi^2 + \sigma^2) \right]
+ i\lambda (\bpi^2 + \sigma^2) \, .
\label{eq:lagrangian_grand}
\end{equation}
The auxiliary field $\lambda$ over which we must integrate to get the
partition function must lack a zero frequency and momentum component
\begin{equation}
\int d^3x d\tau \, \lambda(x) = 0  \, ,
\label{eq:b_dc_less}
\end{equation}
otherwise $\bpi^2(x){+}\sigma^2(x)$ will be identically zero.
The Lagrange multiplier $m^2$ together with $\lambda$ enforces
the averaged constraint
\begin{equation}
\langle \bpi^2 \rangle + \langle \sigma^2 \rangle = f_{\pi}^2 \, .
\label{eq:constraint_grand}
\end{equation}

The next step is to obtain an effective action involving
only the auxiliary field $\lambda$ by integrating out the $\bpi$ and
$\sigma$ degrees of freedom.  At a first glance, this seems like a simple task
since the above Lagrangian (\ref{eq:lagrangian_grand}) is bilinear in
$\bpi$ and $\sigma$.  One is tempted just to carry out the necessary
Gaussian integrals in the partition function
\begin{equation}
Z = \int [d\bpi][d\sigma][d\lambda]\, \exp\left( \int d^3x\,d\tau {\cal L}
\right)
\;.
\end{equation}
However, doing so gives a wrong result.  A Gaussian integral
\begin{equation}
\int d^n x\, \exp\left[ -x_T A x \right]
 = \pi^{n/2}\,{\rm Det}^{-1/2}(A)  \, ,
\end{equation}
is valid only if the determinant of $A$ is positive definite.
When $m^2 = 0$, the action is {\em not} positive definite since
$\displaystyle \int d^3x d\tau \,{\cal L}$ vanishes
when both $\bpi$ and $\sigma$ are constant.
Hence, zero modes exist when $m^2 = 0$,
and if the partition function is to be well-defined
the integration must not be done over constant field configurations.

To carry out the integration over only the non-zero modes
it is convenient to separate the constant zero modes
explicitly from the non-zero ones.
Due to the $O(N)$ symmetry one can always set
$\langle \bpi \rangle = 0$ and $\langle \sigma \rangle = v$.
Shifting $\sigma$ by $v$ then yields
\begin{equation}
{\cal L} =  - \frac{1}{2}\left[ (\partial_{\mu}\bpi)^2 +
(\partial_{\mu} \sigma)^2 + m^2 (\bpi^2 + \sigma^2
+ 2v\sigma) \right]
+ i\lambda (\bpi^2 + \sigma^2 + 2v\sigma)
\label{eq:v_shifted_lagrangian}
\end{equation}
where $\displaystyle \int d^3x d\tau\,\lambda = 0$
is used and a constant term $-m^2 v^2/2$ is discarded.
The condensate $v$ is determined by the condition
\begin{equation}
\langle \sigma \rangle = 0  \, .
\label{eq:condition_v}
\end{equation}

To satisfy the condition (\ref{eq:condition_v}), the action must
not contain terms odd in $\sigma$.
Since $\lambda$ has no $k = 0$ component, the only such term
in the Lagrangian (\ref{eq:v_shifted_lagrangian})
is $-m^2 v \sigma$.
The condition then implies that $m^2$ and $v$ cannot both be nonzero.
As explained in \cite{BK}, there are two possibilities.\footnote
{There is, of course, the possibility that both $m^2$ and $v$ are
zero. However, if this is true, we wouldn't be able to satisfy
both of the conditions (\protect\ref{eq:constraint_grand}) and
(\protect\ref{eq:condition_v}).}

 \begin{enumerate}
 \item $m^2 = 0$ and $v \ne 0$.
 This implies that the symmetry is broken and there are massless
 Goldstone modes.  Since the symmetry is broken at zero temperature,
 this must correspond to low temperatures.

 \item $m^2 \ne 0$ and $v = 0$.
 This implies that the symmetry is restored, and hence there are no
 massless Goldstone modes.  This must correspond to high temperatures.
 \end{enumerate}
The constraint (\ref{eq:constraint_grand}) and the condition
(\ref{eq:condition_v}) must be imposed {\em after} the partition
function is calculated with arbitrary $m^2$ and $v$.

Carrying out the Gaussian integrations over $\bpi$ and $\sigma$
requires completing the square.  In order to do so,
$\sigma$ is shifted once more by a solution of the equation of motion
\begin{equation}
\sigma_\lambda(x) = 2iv\int dz \, G_\lambda(x,z)\, \lambda (z) \, ,
\label{eq:sigma_lambda}
\end{equation}
where
\begin{equation}
G_\lambda \equiv 1/(-\partial^2 + m^2 - 2i\lambda) \, .
\end{equation}
Replacing $\sigma$ with $\sigma {+} \sigma_\lambda$ gives the Lagrangian
\begin{equation}
{\cal L}_{\rm eff} =  -2 v^2 \lambda \, G_\lambda \,\lambda
+ {1\over 2}\bpi{\cdot}(\partial^2 - m^2 + 2i\lambda)\, \bpi
+ {1\over 2}\sigma\,(\partial^2 - m^2 + 2i\lambda)\,\sigma  \, .
\label{eq:L_eff}
\end{equation}
Performing the Gaussian integration over the non-zero modes of
$\bpi$ and $\sigma$ expresses the partition function in
terms of only the auxiliary field $\lambda$.
\begin{eqnarray}
Z = \exp\left\{ -(N/2){\rm Tr}'\, \ln\left(-\partial^2+m^2\right)
\right\} \, \int [d\lambda]\, \exp\left( {\cal S}_\lambda \right) \, ,
\label{eq:partition}
\end{eqnarray}
where
\begin{equation}
{\cal S}_\lambda = -2 v^2\int d^3x d\tau\, \lambda \, G_\lambda \,\lambda
 - (N/2){\rm Tr}\,\ln\left(1 + {2i\lambda \over \partial^2-m^2} \right) \, ,
\label{eq:S_lambda}
\end{equation}
is the effective action for the $\lambda$ field.
The prime on the trace indicates that the zero eigenvalue component
is missing from the operation.
Note, however, that trace-log of
$\left[ 1 + 2i\lambda /(\partial^2 - m^2) \right]$ is well-defined even when
$m^2=0$ since the $\lambda$ field has no $k=0$ component.
Consequently the trace operation in the effective action ${\cal S}_\lambda$
can include the $k=0$ component without any error.

At finite temperature, as argued in \cite{BK}, the residue of the pion
pole in the retarded response function for the axial vector current
is proportional to the square of the pion decay constant.
Hence, calculating the spectral density (see, for example, \cite{Brown})
\begin{equation}
2\pi\,\rho_{\mu\nu}^{a b}(k) \equiv  \int d^3x d\tau
\,e^{-ik\cdot x}\,
\langle [ {\cal A}^a_{\mu}(x), {\cal A}^b_{\nu}(0) ] \rangle
\end{equation}
in the zero momentum and small frequency limit yields $f_\pi^2(T)$.
We now calculate the spectral density by taking the
discontinuity of the Euclidean Green function.  Then, from the
coefficient of the massless pion pole, $f_\pi^2(T)$ is obtained.

As before, one first shifts the sigma field twice to yield
\begin{eqnarray}
{\cal A}^a_{\mu}(x)\, {\cal A}^b_{\nu}(y) & = &
\bigg\{ \partial_{\mu}^x \pi_a(x)
[ v + \sigma(x) + \sigma_\lambda(x) ]
- \pi _{a}(x)
[ \partial_{\mu}^x \sigma(x) + \partial_{\mu}^x \sigma_\lambda(x) ]
\bigg\} \, \nonumber\\
& \times &
\bigg\{ \partial_{\nu}^y \pi_b(y)
[ v + \sigma(y) + \sigma_\lambda(y) ]
- \pi_{b}(y)
[ \partial_{\nu}^y \sigma(y) + \partial_{\nu}^y \sigma_\lambda(y) ]
\bigg\} \, .
\label{eq:AA}
\end{eqnarray}
This expression is to be averaged with the effective Lagrangian given
in eq.~(\ref{eq:L_eff}).
Since the latter is quadratic in $\bpi$ and $\sigma$,
Wick's theorem trivially works with the propagator
\begin{equation}
\langle \pi_a(x)\, \pi_b(y) \rangle
=\langle \sigma(x)\, \sigma(y) \rangle \delta_{ab}
= \langle G_\lambda(x,y) \rangle \delta_{ab}
\, .
\label{eq:prop}
\end{equation}
Performing the $\bpi$ and $\sigma$ integrations is equivalent
to replacing each bilinear in eq.~(\ref{eq:AA}) with $G_\lambda$ and yields
\begin{equation}
\langle {\cal A}^a_{\mu}(x)\, {\cal A}^b_{\nu} (y)
\rangle = {v^2}\,\partial_{\mu}^x \partial_{\nu}^y\,
\langle G_\lambda(x,y) \rangle\delta_{ab} +
\langle  A_{\mu\nu}(x,y) + B_{\mu\nu}(x,y) + C_{\mu\nu}(x,y)
\rangle\delta_{ab} \, ,
\label{eq:after_pi}
\end{equation}
where
\begin{eqnarray}
A_{\mu\nu}(x,y) &=&
2 \big[ G_\lambda(x,y)\, \partial_{\mu}^x \partial_{\nu}^y G_\lambda(x,y)
- \partial_{\mu}^x G_\lambda(x,y)\, \partial_{\nu}^y G_\lambda(x,y) \big] \, ,
\label{eq:A_def}\\
\noalign{\hbox{\vspace{-0.2cm}}}
B_{\mu\nu}(x,y) &=&
v \big[ \partial_\mu^x \partial_\nu^y G_\lambda(x,y)\, \sigma_\lambda(x)
+ \partial_\mu^x \partial_\nu^y G_\lambda(x,y)\, \sigma_\lambda(y)
\nonumber \\
& - & \partial_\mu^x G_\lambda(x,y)\, \partial_\nu^y \sigma_\lambda(y)
- \partial_\nu^y G_\lambda(x,y)\, \partial_\mu^x \sigma_\lambda(x) \big] \, ,
\label{eq:B_def}\\
\noalign{\hbox{\vspace{-0.2cm}}}
C_{\mu\nu}(x,y) &=&
\partial_{\mu}^x \partial_{\nu}^y G_\lambda(x,y)\, \sigma_\lambda(x)\,
\sigma_\lambda(y)
+ G_\lambda(x,y)\, \partial_{\mu}^x \sigma_\lambda(x)\, \partial_\nu^y
\sigma_\lambda(y)
\nonumber  \\
&-& \partial_{\nu}^y G_\lambda(x,y)\, \partial_{\mu}^x \sigma_\lambda(x)\,
\sigma_\lambda(y)
- \partial_{\mu}^x G_\lambda(x,y)\, \sigma_\lambda(x)\, \partial_\nu^y
\sigma_\lambda(y)
\, .
\label{eq:C_def}
\end{eqnarray}
Here the facts
$\langle\bpi\rangle = \langle\sigma\rangle = \langle\sigma_\lambda\rangle=0$
are used to simplify the result.
The expression (\ref{eq:after_pi})
can be now integrated over the remaining $\lambda$ field
with the effective action (\ref{eq:S_lambda}) to yield the axial vector
current correlation function.

First, consider the unbroken phase where $v = 0$.
In this phase the condensate is gone and no pion is
present in the system.  Consequently the pion decay constant must be
zero.

The expression (\ref{eq:after_pi}) is exact.
To use it to calculate $f_\pi^2(T)$ in the
broken phase where $v \ne 0$ and $m^2 = 0$,
one must now distinguish temperatures much lower than
the transition temperature $T_c$ and temperatures close to $T_c$.

First, consider temperatures near $T_c$ so that $v^2(T) \ll NT^2$.
At these temperatures the effective action (\ref{eq:S_lambda})
is of ${\cal O}(N)$.  Hence, the propagator for the $\lambda$ field is
of ${\cal O}(1/N)$.  Ignoring $\lambda$ field contribution all together
in eq.~(\ref{eq:after_pi}) yields for the $(00)$ component of the
Euclidean Green function
\begin{equation}
G_{00}^{ab}(k) = k_0 k_0 v^2 G(k) \, \delta_{ab} +
2 T\sum_{l_0} \int \frac{d^3l}{(2\pi)^3}  l_0 (2 l_0 -  k_0)\,
G(l)\, G(k-l) \, \delta_{ab} \, ,
\label{eq:G_00}
\end{equation}
where $G(k) = 1/k^2$.  The first term in eq.~(\ref{eq:G_00}) clearly
contains a pion pole.  The second term remains finite as
$k\to 0$ after a suitable zero temperature renormalization.
Consequently only the first term contributes to $f_\pi^2(T)$.

To obtain the pion decay constant at finite temperature
one first takes the discontinuity across the real $k_0$ axis
and then divides the result by $k_0^2$.  This results in
\begin{equation}
f_\pi^2(T) = v^2(T) = f_\pi^2 - {N+2\over 12}T^2
\, ; \;\;\;\;\; T \rightarrow T_c \, .
\label{eq:f_pi_near_Tc}
\end{equation}
Here the value of $v^2(T)$ near $T_c$ is extracted from \cite{BK}.
There are ${\cal O}(v^2/N)$ corrections to this result from
further expansions of $v^2 \partial_\mu\partial_\nu G_\lambda$,
$B_{\mu\nu}$ and $C_{\mu\nu}$ terms in eq.~(\ref{eq:after_pi}).
There is also an ${\cal O}(T^2/N)$ correction from further
expansion of $A_{\mu\nu}$.

The behavior of the pion decay constant at low temperature was
first obtained by Gasser and Leutwyler for QCD \cite{GL} and
subsequently verified using a different method by Eletsky
and Kogan \cite{EK}.  For the $O(N)$ sigma model the result
is\footnote{The group $SU(N_f) \times SU(N_f)$ is isomorphic
to $O(N_f^2)$ only for $N_f = 2$ and this limits the quantitative
comparison to two quark flavors.}
\begin{equation}
f_\pi^2(T) = f_\pi^2 - {N-2\over 12}T^2
\, ; \;\;\;\;\; T \ll T_c \, .
\label{eq:f_pi_low_T}
\end{equation}
We have also been able to reproduce this result using the methods
developed in this paper.

To compute the behavior of $f_{\pi}^2(T)$ at arbitrary temperature
is not much more involved than the calculations performed
in this paper.  One just needs to use the full propagator
(\ref{eq:prop}) for the $\lambda$ field instead of the limiting one.
Such a numerical evaluation of $f_\pi^2(T)$ as a function of $T$
would be in itself an interesting project.
However, it is also clear that no surprising feature will emerge from
this calculation since the $\lambda$ propagator is always small compared to
the temperature.  Hence, for most applications that are not
terribly dependent upon fine structure, the following Pad\'{e}
approximation between the
low temperature result (\ref{eq:f_pi_low_T}) and the near $T_c$ result
(\ref{eq:f_pi_near_Tc}) should suffice.
\begin{equation}
f_\pi(T)^2 / f_\pi^2 \, \approx \, {1 - T^2/T_c^2
\over
1 - [4/(N+2)]\, (T^2/T_c^2)\, (1 - T^2/T_c^2) }
\label{eq:Pade}
\end{equation}
The above Pad\'{e} approximation is plotted in Figure 1 for the
case $N = 4$.

\begin{figure}
\setlength{\unitlength}{1cm}

\begin{picture}(0,0)
\put(3.0, -10.1){$0$}
\put(2.7, -9.7){$0$}
\put(4.95, -10.1){$0.2$}
\put(7.05, -10.1){$0.4$}
\put(8.1, -10.7){$T/T_c$}
\put(9.15, -10.1){$0.6$}
\put(11.25, -10.1){$0.8$}
\put(13.35, -10.1){$1.0$}
\put(2.4, -7.8){$0.2$}
\put(2.4, -5.925){$0.4$}
\put(0.5, -5.0){$f_\pi^2(T)/f_\pi^2$}
\put(2.4, -4.05){$0.6$}
\put(2.4, -2.175){$0.8$}
\put(2.4, -0.3){$1.0$}
\end{picture}

\centerline{
\psfig{figure=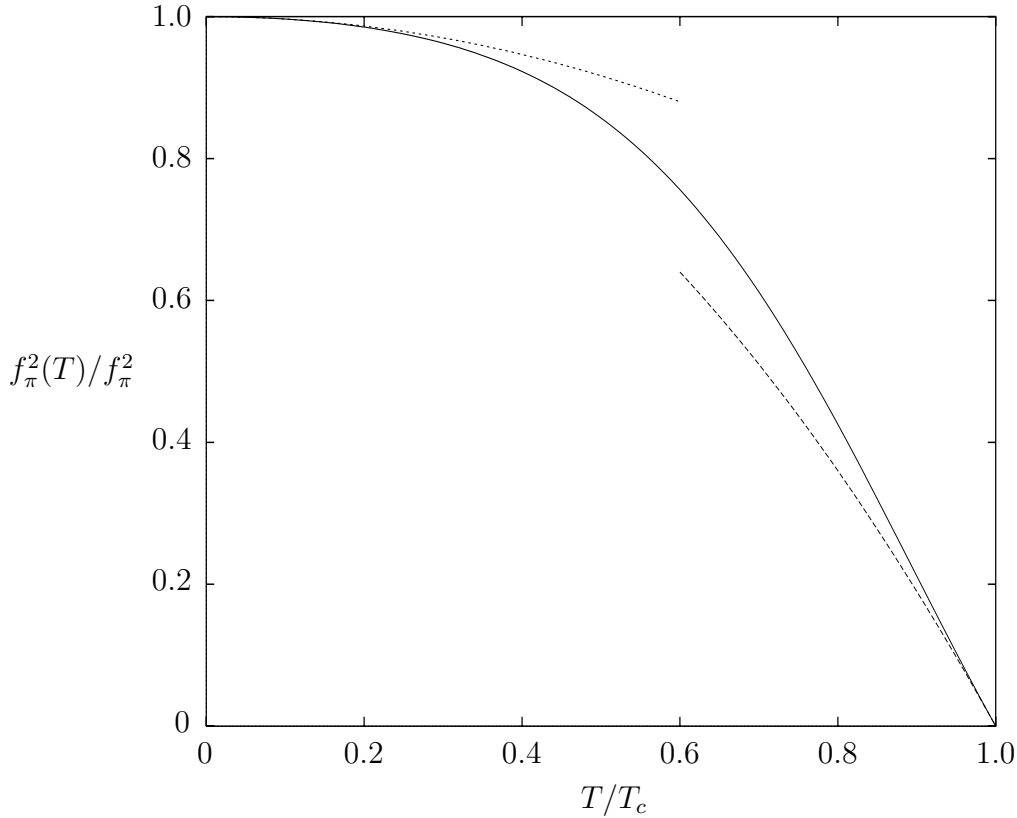,height=10.0cm,width=12.0cm}
}
\vspace{0.7cm}
\label{fig:f_pi_plot}
\caption{\protect\footnotesize Plot of the Pad\'{e} approximant
eq.~(\protect\ref{eq:Pade}) with $N=4$.
The broken lines are limiting behaviors given by
eqs.~(\protect\ref{eq:f_pi_near_Tc}) and
(\protect\ref{eq:f_pi_low_T}).
}
\end{figure}

A natural extension of this work is to include the vector and
axial vector mesons, $\rho, \omega, a_1$, and recompute the
temperature dependence of the pion decay constant.  This would
allow for a more direct and meaningful comparison with lattice
gauge computations.
However, lattice gauge calculations of the pion decay constant
at finite temperature so far have extracted it using the
Gell-Mann Oakes Renner relation \cite{lattice}.  It is not
clear that this is a good definition at finite temperature,
nor whether it is the same as the definition in eq. (4).
In addition, the lattice calculations have used the quenched
approximation which, while of pioneering form, may be missing
some essential physics of the chiral transition.
Further studies in both theoretical and lattice calculation are
clearly called for.

\section*{Acknowledgements}

We thank A. Bochkarev for discussions.  This work was supported by the
U. S. Department of Energy under grant DE-FG02-87ER40328.

\end{document}